\documentclass[aps, prd, amsmath, floats, floatfix, twocolumn,
superscriptaddress, nofootinbib, showpacs]{revtex4-1}
\usepackage[xetex]{graphicx}
\usepackage{color}
\usepackage{soul}
\usepackage{url}
\usepackage{bm}         
\usepackage{times}
\usepackage[normalem]{ulem}

\newcommand{\beq}{\begin{equation}}
\newcommand{\eeq}{\end{equation}}
\newcommand{\beqn}{\begin{eqnarray}}
\newcommand{\eeqn}{\end{eqnarray}}

\newcommand{\lo}{\mathrel{\raise.3ex\hbox{$<$}\mkern-14mu
    \lower0.6ex\hbox{$\sim$}}}
\newcommand{\go}{\mathrel{\raise.3ex\hbox{$>$}\mkern-14mu
    \lower0.6ex\hbox{$\sim$}}}

\newcommand{\RNum}[1]{\uppercase\expandafter{\romannumeral #1\relax}}

\usepackage{color}

\newcommand{\WSU}{\affiliation{Department of Physics \& Astronomy,
	Washington State University, Pullman, Washington 99164, USA}}
\newcommand{\IUCAA}{\affiliation{Inter-University Centre for Astronomy and Astrophysics (IUCAA), 
    Post Bag 4, Ganeshkhind, Pune 411 007, India}}
\newcommand{\CTP}{\affiliation{Center for Theoretical Physics, Polish Academy of Sciences, Al. Lotnikow 32/46, 02-668 Warsaw, Poland}}
\newcommand{\PennState}{\affiliation{Department of Physics, The Pennsylvania State University, University Park, PA 16802, USA}} %
\newcommand{\IGC}{\affiliation{Institute for Gravitation \& the Cosmos, The Pennsylvania State University, University Park, PA 16802, USA}}

\usepackage{graphicx}
\usepackage{dcolumn}
\usepackage{bm}
\usepackage{epsf}

\begin{document}

\title{NONLINEAR MODE-TIDE COUPLING IN COALESCING BINARY NEUTRON STARS WITH RELATIVISTIC CORRECTIONS}

\author{Fatemeh Hossein Nouri} \CTP\IUCAA %
\author{Sukanta Bose} \IUCAA\WSU %
\author{Matthew D. Duez} \WSU %
\author{Abhishek Das} \PennState\IGC %

\begin{abstract}

We compute the 
internal modes
of a non-spinning neutron star and its tidal metric perturbation in general relativity, and determine the effect of relativistic corrections to the modes on 
mode coupling.
Claims have been made that a new hydrodynamic instability can occur in a neutron star in a binary neutron star system triggered by the nonlinear coupling 
of the companion's tidal field to pairs of p-modes and g-modes in it as the binary inspirals toward merger. This ``PG'' instability may be significant since it can influence 
the binary's inspiral phase by extracting orbital energy, thereby potentially causing large deviations in their gravitational waveforms from those predicted by theoretical models that do not account for it. This can result in incorrect parameter estimation, at best, or
mergers going undetected, at worst, owing to the use of deficient waveform models. On the other hand, better modeling of this instability and its effect on binary orbits can unravel a new phenomenon and shed light on stellar instabilities, via gravitational wave observations. 
So far, all  mode-tide coupling instability studies have been 
formulated in Newtonian perturbation theory.  Neutron stars are compact objects, so relativistic corrections might be important. We present and test a new code to calculate
the relativistic eigenmodes of nonrotating relativistic
stars.  We use these relativistic tide and neutron star eigenmodes to compute the mode-tide 
coupling strength (MTCS) for a few selected equations of state.  The MTCS thus calculated can be at most tens of percent different from its purely Newtonian value, but we confirm the dependencies on orbital separation and equation of state found by Newtonian calculations.  For some equations of state, the MTCS is very sensitive to the neutron star crust region, demonstrating the importance of treating this region accurately.
[{\color{red}This manuscript has been assigned the LIGO preprint number LIGO-P2100021.}]

\end{abstract}


\maketitle

\section{Introduction}
\label{sec:intro}

Dozens of compact binary mergers, involving neutron stars (NSs) and black holes (BHs),
have been made using the LIGO~\cite{TheLIGOScientific:2014jea} and Virgo~\cite{TheVirgo:2014hva} gravitational wave (GW) detectors~\cite{LIGOScientific:2020ibl,LIGOScientific:2021qlt}.  These GW signals can be used to probe physics involving relativistic high density matter in strong gravity. One phenomenon proposed to occur in binaries with at least one neutron star, and nearing merger, is the tide-induced instability originating from the coupling of its acoustic modes (or p-modes) and g-modes, where the latter are associated with buoyancy arising from stratification in the NS core. 
While mode-tide couplings, per se, have been studied in numerical models of non-relativistic binaries of main sequence stars~\cite{Weinberg_2012}, which are far less dense than NS matter, nevertheless no claims for the aforementioned tide-induced instabilities have been made for such stars. However, if such instabilities are found to occur in binary neutron stars (BNSs), they will not only provide clues to the NS
equation of state (EOS),
but also have important implications for merger waveforms and possible missed detections if the instability-induced modulations cause significant dephasing not accounted for by GW search algorithms~\cite{Essick_2016}. Such a detection would also have interesting implications on binaries of main sequence stars.

The first studies of couplings between tides and internal modes of neutron stars in a BNS considered the possible importance of resonant coupling, which will occur during inspiral when the orbital period comes into resonance with neutron star g-modes~\cite{Reisenegger_1994,Lai:1993di,Shibata_1994,Wynn_1999,Lai_2006,Flanagan_2007}.  Another possibility that has generated great excitement is one where the three-mode couplings can be strong even away from resonance, when the two daughter modes have similar wave numbers~\cite{Wu_2001,Weinberg_2013,Venumadhav_2013,Essick_2016,Weinberg_2016}.

Attempts to detect this instability via GW phase deviation in the BNSs GW170817 and GW190425 have been made~\cite{Weinberg:2018ic,Abbott:2020uma}. It was found that the observed signal is consistent with the absence of this instability in that system. 
This comparison also constrained the p-g amplitude for $1.4M_\odot$ neutron stars to less than a few tenths of the theoretical maximum, and the p-g saturation frequency to $\sim 70$Hz~\cite{Weinberg:2018ic}. However, these studies, as well as some notable alternative
treatments~\cite{Reyes:2018bee}, did not preclude the possibility of this phenomenon occurring in the inspiral of other binaries involving neutron stars.

Studies of the non-resonant coupling to date have been non-relativistic.  The partial exception is the work of Zhou and Zhang~\cite{Zhou_2017}, who used Newtonian perturbation theory on relativistic (Oppenheimer-Volkoff) neutron star equilibria to study the static and dynamic coupling between tides and g-modes.  Using six plausible neutron star equations of state, they find significant dependence of the instability growth rate primarily on the Brunt-V\"{a}is\"{a}l\"{a} frequency and secondarily on the neutron star compaction.  However, hybrid approaches of this sort are not necessarily more accurate than purely Newtonian models, and in fact the inconsistency between equilibrium and perturbation physics can sometimes lead to distinct errors of its own (see, e.g.,~\cite{Reisenegger_1994b}).
A consistently relativistic treatment is needed.

In this paper, we take a
step toward this goal by treating equilibria, tidal responses -- both static and dynamic, and g-mode eigenfunctions fully relativistically.  Only the tidal coupling integral remains non-relativistic.  Because we focus only on the effect of relativity, we use the common idealized treatment of neutron star microphysics, ignoring possible effects of the neutron star crust or superfluidity.  We do, however, use several newer nuclear physics equations of state.
We closely follow the same approach as the one pursued in Z\&Z; i.e., we apply the relativistic corrections only to the MTCS term, which is linear in the tidal strength $\epsilon$ for the shifted modal frequency. 
This term includes only the coupling between the tide and the g-modes. Therefore, it can not be considered as a complete calculation of the p-g instability,
although
it does involve low-frequency higher order g modes that contribute to it.
In any case, the g mode - tide coupling may be more important than the p-g instability~\cite{Venumadhav_2013}.

This paper is organized as follows. In Sec.~\ref{sec:formalism} we introduce our relativistic corrections, and Tolman-Oppenheimer-Volkoff (TOV) star setups with a brief comparison between EOSs. 
The results and discussion are presented in Sec.~\ref{sec:results}. 
Section~\ref{sec:conclusion} summarizes our conclusions. In  Appendix~\ref{sec:app_methods} we explain our numerical methods and present various test results that were used to examine our code.
The importance of  relativistic corrections to higher-order modes coupling is discussed briefly in Appendix~\ref{sec:app_higher_order}.

We use units such that $G = c = M_{\odot} = 1$, unless otherwise specified. We convert to physical units when calculating observable quantities.

\section{Star Setups and Relativistic Corrections}
\label{sec:formalism}

In this paper our goal is to compute the mode-tide coupling strength (MTCS) in a BNS system. 
This problem was originally studied
by Weinberg et al.~(2012)~\cite{Weinberg_2013}
and Venumadhav et al. (2013)~\cite{Venumadhav_2013}. 
Here we follow the same Newtonian approach to the mode coupling as that taken by Zhou and Zhang (2018) (Z\&Z)~\cite{Zhou_2017}, mainly computing the MTCS using Eq.~(33) for static tides
and Eqs.~(43)-(45) for non-static tides from Z\&Z,
with new modifications to consider certain relativistic effects. Specifically,
(1) the eigenfrequencies and eigenfunctions of the stellar modes are computed relativistically; 
(2) the dynamical tidal displacement due to the companion star is computed in the relativistic formalism; 
(3) we employ all hydrodynamic variables 
in a relativistically 
consistent way. This work is limited to the 
special case where the masses of both stars are identical.

Our motivation is to study the significance of the relativistic corrections. Therefore, we limit MTCS computations to a single case of g-mode with $n=32$, $l_g=4$ (i.e.,
a ``g-mode, g-mode, tide coupling" or, equivalently, a coupling of two g modes and the tide)
for studying its sensitivity to the EOS. 
We mostly focus on the dynamical tides since the nonlinear mode-tide coupling is stronger by up to a few orders of magnitude compared to static tides and, hence, are expected to be more important for stability and gravitational-wave observations (see Sec.~4.4 in Z\&Z).
We also compute a few cases of static tide MTCS in order to test our results against Z\&Z.

\subsection{TOV Star and Equations of State}

We set up the background star by solving the TOV equations (see, e.g., Eq.~(5) in Z\&Z). For each EOS, the central density is chosen to set the total gravitational mass to $1.4 M_{\odot}$.
The radii and central densities of TOV stars with different EOS are shown in table~\ref{tab:EOS}.

The unperturbed metric is static and spherically symmetric and, therefore, may be written in the form
\begin{equation}
    ds^2 = -e^{\nu}dt^2+e^{\lambda}dr^2+r^2(d\theta^2+\sin^2\theta d\phi^2),
\end{equation}
where the metric components are related to the fluid density $\rho(r)$ and pressure $p(r)$ via Einstein's equations 
%
\begin{equation}
\frac{d\nu}{dr} = 2\frac{(m(r)+4\pi r^3p)}{r(r-2m(r))}\,,
\end{equation}
\begin{equation}
e^{-\lambda} = 1-\frac{2m(r)}{r}\,,
\end{equation}
and $m(r)$ is the mass interior to radius $r$
%
\begin{equation}\label{eq:m-of-r}
m(r) = \int_{0}^{r}{4\pi r^2 \rho dr}\,.
\end{equation}
We use three finite-temperature, composition-dependent nuclear-theory based equations of state, all derived with relativistic mean field (RMF) description of the nuclear matter. They are publicly available in fairly high-resolution tabulated form~\cite{OConnor:2010} at~\cite{stellarcollapse}.
The electron fraction $Y_e$ of each equilibrium star is determined by the beta-equilibrium condition. 

1- Shen~\cite{Shen_1998}: An RMF EOS that covers broad density and temperature ranges ($10^5 < \rho < 10^{15.5} \mathrm{g/cm^3}$; $0 < T < 100$MeV). 
For our purpose, we extract the table for the minimum temperature, i.e., $T=0.1$. Shen is our stiffest EOS, giving the radius of a  $1.4M_{\odot}$ neutron star to be $R_{1.4M_{\odot}}=14.53$km.

2- DD2~\cite{Hempel_2012}: An RMF with a density dependent nucleon-meson coupling.
DD2 has intermediate compactness in our selection, and gives $R_{1.4M_{\odot}}=13.22$km.

3- SFHo~\cite{Steiner_2013}: Another RMF using a covariant Walecka model Lagrangian (ensuring causal sound speeds). This is the softest EOS in our selection, which makes $R_{1.4M_{\odot}}=11.9$km.

While the above selection of EOSs is not exhaustive, nevertheless it includes EOSs that are of varying stiffness and are allowed by both the GW170817 observations~\cite{Abbott:2018exr,Biswas:2020puz} and the maximum mass constraint arising from pulsar observations~\cite{Cromartie:2019kug}. 

Following the same criterion adopted by Z\&Z and Lai (1994)~\cite{Lai:1993di}, the crust-core boundary is taken to be the radius 
where the electron fraction
$Y_e$ is at its minimum. This criterion helps us to distinguish the core g-modes from their crust counterparts. Furthermore, it is used to cut off g-modes by setting the Brunt-V\"{a}is\"{a}l\"{a} frequency ($N$) to zero outside of this radius.
The electron fraction versus density is shown in Fig.~\ref{fig:Ye-EOS} for the EOS selection. 

\begin{figure}
  \includegraphics[width=\linewidth]{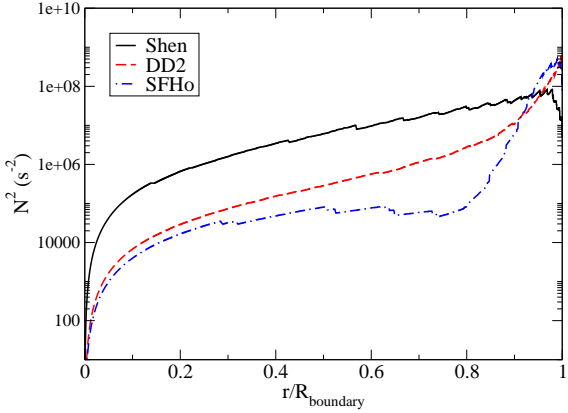}
  \caption[N2-EOS]{
The Brunt-V\"{a}is\"{a}l\"{a} frequencies plotted as functions of the star's radial coordinate $r$ for Shen, DD2 and SFHo equations of state.}
  \label{fig:N2-EOS}
\end{figure}

\begin{figure}
  \includegraphics[width=\linewidth]{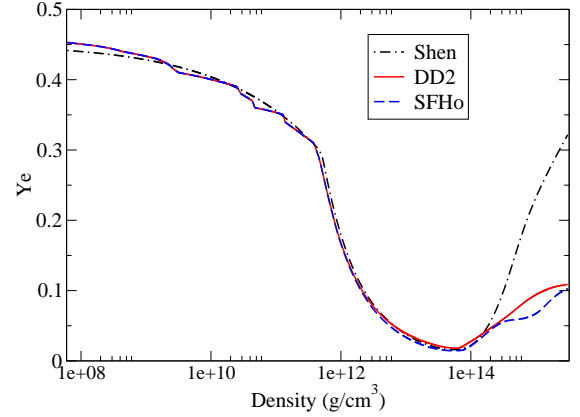}
  \caption[Ye-EOS]{
  The electron fraction versus density for Shen, DD2 and SFHo.
}
  \label{fig:Ye-EOS}
\end{figure}

According to Z\&Z, the Brunt-V\"{a}is\"{a}l\"{a} frequency plays an important role in determining the strength of the nonlinear coupling between g-modes and tides. This will be discussed in more detail in the results' section. Figure~\ref{fig:N2-EOS} compares the buoyancy frequency of different equations of state,
computed with Eq.~(27) from Z\&Z.  When computing this frequency, we take numerical derivatives in the EOS table, which may produce errors of a few percent, but is adequate for our purposes.
For much of the star's body, these frequencies track its compactness, with Shen as the highest, SFHo as the lowest, and DD2 in the middle. 

\begin{table}
\begin{ruledtabular}
\begin{tabular}{ l l l l }
EOS & $\rho_{0c}~$ $\mathrm(g/cm^3)$ & $R$ (km) & $R_\text{boundary}$ (km) \\
\hline                                         
Shen & $5.0 \times 10^{14}$ & 14.532 & 13.453 \\
DD2  & $5.8 \times 10^{14}$ & 13.220 & 12.297 \\ 
SFHo & $8.5 \times 10^{14}$ & 11.901 & 11.174 \\
\end{tabular}
\end{ruledtabular}
\caption{\label{tab:EOS}
The rest-mass central density $\rho_{0c}$, radius of TOV stars $R$, and core-crust boundary radius $R_b$ for Shen, DD2,
and SFHo equations of state are tabulated above. The central density has been adjusted so that the star's mass always equals $1.4 M_{\odot}$.\\
}
\end{table}

\subsection{Relativistic Perturbation Equations}

We follow the standard approach laid out in Thorne and Campolattaro~\cite{Thorne_1967} for relativistic perturbation equations. By adopting the notation of Lindblom and Detweiler (1983)~\cite{Lindblom:1983ps}, the perturbed metric tensor is described as:
\begin{eqnarray}
\label{eq:per-metric}
ds^2 &=& -e^{\nu}(1+r^lH_0Y_{lm} e^{i\omega t})dt^2 \nonumber\\
&&- 2i\omega r^{l+1}H_1Y_{lm} e^{i\omega t} dtdr \nonumber\\
&&+ e^{\lambda}(1-r^l H_0 Y_{lm} e^{i\omega t})dr^2 \nonumber\\
&&+ r^2(1-r^lKY_{lm} e^{i\omega t})(d\theta^2+\sin^2 \theta d\phi^2)\,,
\end{eqnarray}
where the radial dependence of the metric perturbations is characterized by $H_0$, $H_1$ and $K$ -- all functions of the radial coordinate $r$.  
The angular-coordinate dependence of the perturbations is represented by spherical harmonics $Y_{lm}$. Also, the time dependence $e^{i\omega t}$ of the perturbations is characterized by mode angular frequency $\omega$.

The Lagrangian fluid displacement vector components in a perturbed star are~\cite{Lindblom_1990}:
\begin{equation}\label{eq:xi-r}
\xi^r = r^{l-1}e^{-\lambda/2}WY_{lm}e^{i\omega t} \,,
\end{equation}
\begin{equation}\label{eq:xi-theta}
\xi^\theta=-r^{l-2}V\frac{dY_{lm}}{d\theta}e^{i\omega t}\,,
\end{equation}
\begin{equation}\label{eq:xi-phi}
\xi^\phi=-\frac{r^l}{r^2 \sin^2\theta}V\frac{dY_{lm}}{d\phi}e^{i\omega t}\,,
\end{equation}
where $W$ and $V$ are the two fluid perturbation functions.

The perturbation functions are related as a consequence of Einstein's equations (see Eq.~(5) from ~\cite{Detweiler_1985}):
\begin{eqnarray}
\label{eq:H0}
\left[3m(r)+\frac{1}{2}(l+2)(l-1)r+4\pi r^3 p \right]H_0 =\quad\quad \quad\quad\quad\quad &&  \nonumber \\
8\pi r^3 e^{-\nu/2}\chi - \left[\frac{1}{2}l(l+1)(M+4\pi r^3p) - \omega^2 r^3 e^{-(\lambda + \nu)}\right]H_1 && \nonumber \\
+ \Big[\frac{1}{2}(l+2)(l-1)r - \omega^2 r^3 e^{-\nu} 
\quad\quad\quad\quad && \nonumber \\
- r^{-1}e^{\lambda}(m(r)+4\pi r^3 p)(3m(r)-r+4\pi r^3 p)\Big]K\,, \quad\quad&& 
\end{eqnarray}
where the perturbation function $\chi$ is defined as
\begin{eqnarray}
\label{eq:X-def}
\chi \equiv &~~& \omega^2 (\rho+p)e^{-\nu/2}V -r^{-1}p^{\prime} e^{(\nu - \lambda)/2}W \nonumber \\
&& + \frac{1}{2}(\rho+p)e^{\nu/2}H_0
\end{eqnarray}
and $\rho$ and $p$ are energy density and pressure of the unperturbed star, respectively. A prime denotes the derivative with respect to $r$, and $m(r)$ is the local mass as defined by Eq.~(\ref{eq:m-of-r}).

\subsection{Relativistic Tidal Displacement Computation}
\label{subsec:tide_dis}

The tidal displacement represents the response of a 
fluid star to the external non-static tidal field, which
couples nonlinearly to the normal modes of neutron star and results in frequency shift.
The Newtonian tidal displacement is computed by solving a system of four ODEs 
representing the Newtonian perturbation equations including the tidal field of the companion star
as the source terms (see Eqs.~(C1) from Z\&Z, and Eqs.~(A9-A11) from Weinberg et al. (2012)~\cite{Weinberg_2012}).

In the relativistic approach, we need to compute the perturbation of the metric and hydrodynamic functions including the effects of the tidal field of the companion star. 
For a relativistic perturbed star, the interior solution is derived by solving a system of four coupled ODEs, namely:
\begin{equation}\label{eq:H1}
\begin{aligned}
{H_1}^\prime = -r^{-1}(l+1+2Me^{\lambda}r^{-1}
+ 4\pi r^2 e^{\lambda}(p-\rho))H_1 \\
+ r^{-1} e^{\lambda} (H_0 + K - 16\pi (\rho + p)V), \\
\end{aligned}
\end{equation}

\begin{equation}\label{eq:K}
\begin{aligned}
{K}^\prime = r^{-1}H_0 + \frac{1}{2}l(l+1)r^{-1}H_1 - ((l+1)r^{-1} 
- \frac{1}{2} {\nu}^\prime)K \\
-8\pi(\rho + p)e^{\lambda/2}r^{-1}W,
\end{aligned}
\end{equation}

\begin{equation}\label{eq:W}
\begin{aligned}
{W}^\prime = -(l+1)r^{-1}W + re^{\lambda/2}(\gamma^{-1}p^{-1}e^{-\nu/2}\chi \\ 
- l(l+1)r^{-2}V + \frac{1}{2}H_0 +K),
\end{aligned}
\end{equation}

\begin{equation}\label{eq:X}
\begin{aligned}
{\chi}^\prime = -lr^{-1}\chi + (\rho+p)e^{\nu/2}\Big\{\frac{1}{2}(r^{-1}-\frac{1}{2}\nu^{\prime})H_0 \\
+\frac{1}{2}\big[ r\omega^2 e^{-\nu} + \frac{1}{2}l(l+1)r^{-1})H_1 \\
+ \frac{1}{2}(\frac{3}{2}\nu^\prime - r^{-1})K - \frac{1}{2}l(l+1)\nu^{\prime}r^{-2}V - r^{-1}(4\pi(\rho+p)e^{\lambda/2} \\
+ \omega^2 e^{\lambda/2-\nu} -\frac{1}{2}r^{2}(r^{-2}e^{\lambda/2}\nu^{\prime})^{\prime}\big]W\Big\}\,,
\end{aligned}
\end{equation}
where $\gamma$ is the adiabatic index.
We integrate Eqs.~(\ref{eq:H1}) and ~(\ref{eq:K}) for $H_1$ and $K$ to solve for the metric perturbation, and Eqs.~(\ref{eq:W}) and ~(\ref{eq:X}) for $W$ and $\chi$ to obtain the hydrodynamic perturbation.  Algebraic constraints (Eqs.~(\ref{eq:H0}) and ~(\ref{eq:X-def})) give $V$ and $H_0$.

The solutions must be regular at the origin have vanishing perturbed pressure at the surface ($\chi=0$).  The values for the metric perturbation functions $H_1$ and $K$ at the surface are given by the exterior solution described below. 

The oscillations of the fluid and the metric are assumed to be synchronous with the dynamical tidal field, 
which means that $\omega$ depends on the  angular frequency of the tidal field $\Omega$ through the relation $\omega = m\Omega$, where $m=2$ for the dynamical tide.
$\Omega$ itself deviates from the orbital angular frequency in Kepler's law by higher-order post-Newtonian terms, as can be seen in Eq.~(9.25) of Ref.~\cite{Poisson:2018qqd}.  
We evolve the perturbation equations
both outward from the origin $r=0$ and inward from the surface $r=R$, and use a shooting method to match the solutions at a fitting radius $r=R/2$. The details of our numerical scheme are given in the appendix~\ref{sec:app_methods}.

For the exterior solution, we solve the Zerilli equation~\cite{Zerilli_1970},
\begin{equation} \label{eq:zerilli}
\begin{aligned}
\left(\frac{d^2}{dr^*}+\omega^2-V(r)\right)Z=0
\end{aligned}
\end{equation}
with the potential given by
\begin{equation}
\begin{aligned}
V(r) = \frac{2(r-2M)}{r^4(\varsigma r + 3M)^2}[\varsigma^2(\varsigma+1)r^3 + \\
3\varsigma^2Mr^2 + 9\varsigma M^2 r +9M^3]\,,
\end{aligned}
\end{equation}
where $\varsigma \equiv (l-1)(l+2)/2$. We do so by matching the Zerilli function to the asymptotic solution at large radii.
The asymptotic solution is taken to be the dynamical tidal field from an orbiting mass around a Schwarzschild black hole, calculated by Fang and Lovelace (2005)~\cite{Fang_2005}.
\begin{equation}
\begin{aligned}
Z = B \left(1+\frac{4i\eta}{3}\right)
\Big[\tilde{r}^3 + \frac{3\tilde{r}^2}{4}
- \frac{9\tilde{r}}{16} - \frac{21}{64} + \frac{63}{256\tilde{r}}\Big] \\
+ B\Big[\frac{-945-236i\eta}{5120\tilde{r}^2} + \frac{8505+15436i\eta}{61440\tilde{r}^3}+O\left(\frac{1}{\tilde{r}^4}
\right)\Big]\,,
\end{aligned}
\end{equation}
where $B$ is defined as 
\begin{equation}
\begin{aligned}
B = \frac{4 M^4}{A^3} \sqrt{\frac{2\pi}{15}}
\left(1-\frac{4i\eta}{3}\right).
\end{aligned}
\end{equation}
In these equations, $\eta = 2M\omega$, $\tilde{r}=r/2M$,
and $r^*$ is the tortoise coordinate $r^* = r+2M \log(r/2M-1)$, and $A$ is the binary separation.

The relation between the metric perturbation functions ($H_1$,$K$) and the Zerilli function 
are given by equations (B1-B2) of Ref.~\cite{Fang_2005},
\begin{equation}
\begin{aligned}
H_1 = -i\omega \frac{\varsigma r^2 -3\varsigma Mr - 3M^2}{r^2(r-2M)(\varsigma r + 3M)}Z - \frac{i\omega}{r} \frac{dZ}{dr}\,,
\end{aligned}
\end{equation}

\begin{equation}
\begin{aligned}
K = \frac{\varsigma (\varsigma+1)r^2+3\varsigma Mr + 6M^2}{r^4(\varsigma r +3M)}Z +\frac{1}{r^2} \frac{dZ}{dr^*}.
\end{aligned}
\end{equation}
Note that these equations are different from their counterparts in Ref.~\cite{Fang_2005} -- by factor of $r^l$ -- to adopt them to the notation of Lindblom and Detweiler~\cite{Lindblom:1983ps} used for the interior solution. 

\begin{figure}
  \includegraphics[width=\linewidth]{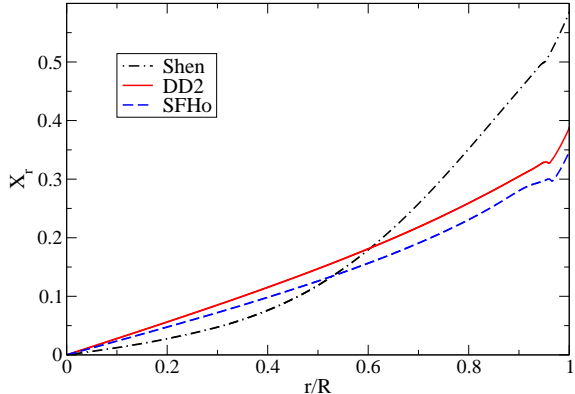}
  \caption[Rel-TideDis]{
The relativistic tidal displacement for Shen, DD2 and SFHo equations of state computed at $A=70$km binary separation, for $m=2$ non-static tide in a 1.4 $M_\odot$ NS.
The units of $X_r$ here are the same as in Z\&Z,
namely, $\varepsilon R W_{22}$, where $R$ is the radius of the star,
$\varepsilon$ is the tidal strength $R^3/A^3$, and $W_{lm}$ are the coefficients in the expansion of the tidal
potential in terms of spherical harmonics. In particular, $W_{22} = \sqrt{3\pi/10}$. 
    }
  \label{fig:Rel-TideDis}
\end{figure}

\subsection{Estimate of errors due to the test-mass approximation} 

Strictly speaking, the aysmptotic behavior of the perturbed metric solution calculated by Fang and Lovelace is for the case where the perturbing object is an orbiting secondary with mass $\ll M$. In comparison, the mass of the secondary in our binary is the same as that of the primary.
While the two problems are not identical, the asymptotics of the latter problem can be mapped to the first while incurring an error in the tidal displacement of about 20\% or less for a neutron star mass under $2M_\odot$ and a binary separation of over 50~km. One way to check this assertion is to compare the Fang and Lovelace (FL) asymptotics with those of the comparable-mass binary studied by Poisson and Corrigan (PC)~\cite{Poisson:2018qqd}, who computed the first order post-Newtonian (1PN) correction terms of the tidally perturbed metric of the primary. Both works study the perturbed metric of the compact object of interest at a spacetime point located at an areal radius $b$ such that 
$M \ll b \ll A$. 
In this limit, the metric of FL (which takes the companion to be a test mass) deviates from the comparable-mass metric of Ref.~\cite{Poisson:2018qqd} 
by terms at 1PN and beyond. 

The magnitude of the 1PN term found in PC can be estimated for the
type of maximum mass and minimum separation discussed above. For close binary separations MTCS magnitudes computed here will not be trustworthy. All the same, the ensuing inaccurate phase distortions in GWs from the binary will not impact its detectability owing to the short time remaining before binary merger. In the earlier part of the orbital phase the errors are smaller and will impact orbital deviations less.

For estimating the error incurred in the metric arising from approximating the secondary as a test-mass, consider the spacetime point of interest to be somewhat away from the surface of the primary star, of mass $M=2M_\odot$, say, at $b \sim 30$km. Also, the mass of the secondary in our case (for the comparable-mass binary) is $M_2 = M =6$km, and the closest binary separation studied here is $A=50$km. For this scenario, the FL and PC metrics differ by 
$O(b^2 M_2/A^3) \approx 2.5\%$. This can be seen by comparing Eq.~(28) of FL with Eq.~(5.3) of PC. While this departure is small, the main error in our results on the tidal displacement arises from approximating $\Omega$ to be the Keplerian orbital angular frequency. Relative to it, the 1PN correction in the tidal angular frequency can be deduced from Eq.~(9.25) of PC, and is
\begin{equation}
\frac{1}{2}
\left[ 3 + \frac{M M_2}{(M+M_2)^2}
\right]\frac{M^{3/2}}{A^{5/2}} =
\frac{1}{2}\cdot
\frac{13}{4}\cdot
\frac{6}{50}
\sqrt{\frac{M}{A^3}}\,,
\end{equation}
which is $\approx 20
\%$ 
different from the 1PN corrected value for that frequency. This can cause a similar difference in the true value of the tidal displacement.  Like the metric error, this frequency error is much smaller at larger separations.

To understand the nature of our approximations, it is helpful to remember that there are two compactions in a binary neutron star system:  the compaction of the individual stars $M/R$ and the compaction of the binary itself $2M/A$.  Even at large separations, $M/R$ is $O(10^{-1})$, and the system is relativistic.  Our formalism keeps terms that are higher-order in $M/R$ but drops terms that are post-Newtonian in $M/A$, the latter only becoming large in the very late inspiral when the effects we study are not expected to be important.

\subsection{Relativistic Eigenmodes}
\label{subsec:eigenmodes}

In order to compute the MTCS integral one needs to derive the frequencies and Lagrangian displacement vectors of the perturbed star for p-modes and g-modes. For our particular studies, we are interested in high-order modes coupled with the tides.
Following Z\&Z we use the Cowling approximation, which provides enough accuracy at least for high-order mode (high mode number) computations, but in the relativistic formalism.

To derive the internal modes (p-modes and g-modes) within the Cowling approximation,
we solve the perturbation equations for the radial component of the Lagrangian displacement and Eulerian perturbation for pressure. 
In other words, in the relativistic formalism and within the Cowling approximation, we ignore the perturbations of the metric, and the original four coupled ODEs, Eqs.~(\ref{eq:H1}-\ref{eq:X}), reduce to two coupled ODEs considering only hydrodynamic perturbations. 
For this we refer to Lindblom~\&~Splinter~(1990)~\cite{Lindblom_1990} and Finn ~1988)~\cite{Finn_1988}. 

Adopting the same notation as Finn~(1988)~\cite{Finn_1988}, we define the new variables, the radial component of the displacement vector in the orthonormal basis and redshifted pressure perturbation, as follows:
%
\begin{equation}
\hat{\xi^r}=e^{\lambda/2}\xi^r = r^{l-1} W(r) Y_{lm}e^{i\omega t} \,,
\label{eq:xiHat}
\end{equation}
\begin{equation}
\delta P=e^{\nu/2}\delta p \,,
\end{equation}
where $\delta p$ is the Eulerian pressure perturbation. Also, we used Eq.~(\ref{eq:xi-r}) above to replace $\xi^r$. 
Note that the radial dependence of  $\hat{\xi^r}$ 
is captured purely by $X_r\equiv r^{l-1} W(r)$.
This quantity 
was studied by Z\&Z in the Newtonian computation.
In comparison, our Fig.~\ref{fig:Rel-TideDis} shows the tidal displacement, computed relativistically for different EOSs at binary separation $A=70$km.
This confirms the previous investigations for dynamical tides, such as of Maselli et al. (2012)~\cite{Maselli_2015}, that the deformation of a star
due to the external tidal field is larger for a stiffer EOS during late inspiral.
Moreover, in comparison to Z\&Z relativistic effects appear to reduce $X_r$.

In terms of these new variables, and assuming $e^{i\omega t}$ time dependence, the perturbation equations become
\begin{equation}\label{eq:relxi}
\hat{\xi^r}'=\left[-\frac{2}{r}-\frac{p'}{p\gamma}\right]\hat{\xi^r}+\left[\frac{l(l+1)e^\nu}{(\rho+p)\omega^2r^2}-\frac{1}{p\gamma}\right]e^{(\lambda-\nu)/2}\delta P \,,
\end{equation}
\begin{equation}\label{eq:relp}
\begin{aligned}
(\delta P)'=(\rho+p)\Big[e^{(\lambda-\nu)/2}\omega^2+e^{(\nu-\lambda)/2}\frac{p'}{\rho+p}\Big(\frac{p'}{\gamma p} \\
-\frac{\rho'}{\rho+p}\Big)\Big]\hat{\xi^r}+\frac{p'}{p\gamma}\delta P.
\end{aligned}
\end{equation}
The eigenfrequency $\omega$, and eigenfunctions, $\delta P$ and $\hat{\xi^r}$, are derived by solving these equations numerically using the shooting method, and applying the same boundary conditions as in the Newtonian perturbation equations. 
Those conditions require (a) the
Lagrangian pressure perturbation to vanish at the surface and (b) the
regularity condition, $\xi_r=l\xi_h$, to hold at the center, with
$\xi_h$ denoting the component of the Lagrangian displacement vector transverse to the radial direction~\cite{DalsgaardBook}.

\begin{figure}
  \includegraphics[width=\linewidth]{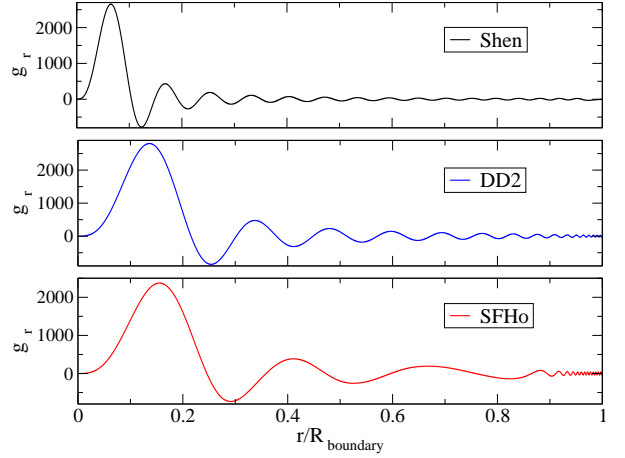}
  \caption[eigenfunc-EOS]{ 
The radial displacement eigenfunction $g_r$ for the $l = 4$, $n = 32$ g-mode
with frequencies
listed as $f_{g(R)}$ in table
\ref{tab:MTCS_results}, computed with our relativistic code in the Cowling approximation, for Shen, DD2 and SFHo EOSs. The eigenvectors are normalized as $\omega_g^2 \int{d^3x\rho | \overrightarrow{g} |^2} = E_0$, where $E_0 = M^2/R$.
    }
  \label{fig:eigenfunc-EOS}
\end{figure}

The relativistic eigenfrequencies and radial component of the displacement vector $\hat{\xi^r}$ derived for g-modes $n=32$, $l=4$ for Shen, SFHo and DD2 equations of states are shown in 
Table~\ref{tab:MTCS_results} and Fig.~\ref{fig:eigenfunc-EOS}.
In the Z\&Z paper, the radial and horizontal components of the Lagrangian displacement vector of g-modes are called $g_r$ and $g_h$ respectively. For convenience, we follow the same convention henceforth. 
The relation between the displacement vector components and the perturbation functions $W$ and $V$ are given in Eqs.~(\ref{eq:xi-r}-\ref{eq:xi-phi}). In the relativistic Cowling formalism the horizontal component of the Lagrangian displacement is related to the Eulerian pressure perturbation as follows:
\begin{equation}
\xi_h = \frac{e^{\nu/2}}{r\omega_g^2}\frac{\delta p}{p+\rho},
\end{equation}
where $\omega_g$ is the g-mode frequency.

\subsection{Mode-tide coupling strength} 
\label{subsec:MTCS-and-pgi}

As is explained in Z\&Z, the g-mode frequency shift due to its nonlinear coupling to a p-mode and the tide can be expressed as:
\begin{equation} \label{eq:freq_shift}
\begin{aligned}
\frac{\omega_{-}^{2}}{\omega_{g}^{2}} = 1 - C_1 - C_2 - C_3 + \mathcal{O}(\epsilon^3)\,,
\end{aligned}
\end{equation}
where
$\omega_g$ is the  g-mode's original frequency and $\omega_-$ is its shifted frequency. The terms $C_1$, $C_2$ and $C_3$ are defined as follows:
\begin{equation} \label{eq:lin_C1}
\begin{aligned}
C_1 = \epsilon \left(U_{\overline{g}g} + \sum_{a} 2 \kappa_{a\overline{g}g} \chi_a ^{(1)} \right)\,,
\end{aligned}
\end{equation}
\begin{equation} \label{eq:nonlin_C2}
\begin{aligned}
C_2 = \epsilon^2 \sum_{a,b} \left( 2 \kappa_{a\overline{g}g} \chi_a ^{(2)} 
+ 3 \kappa_{ab\overline{g}g} \chi_a ^{(1)} \chi_b ^{(1)} \right)\,,
\end{aligned}
\end{equation}
\begin{equation} \label{eq:nonlin_C3}
\begin{aligned}
C_3 = \epsilon^2 \frac{\omega_p^2}{\omega_p^2 - \omega_g^2} \left \lvert U_{\overline{p}g} + \sum_{a} 2 \kappa_{a\overline{p}g} \chi_a ^{(1)} \right \rvert^2\,.
\end{aligned}
\end{equation}  
Here, $\epsilon \equiv R^3/A^3$ is the tidal strength,
$\omega_p$ is the p-mode frequency,
$\chi^{(1)}$ is the response of the neutron star to the tide, $\kappa_{a\overline{p}g}$,  $\kappa_{ab\overline{g}g}$ and $U_{\overline{p}g}$ are the coupling strengths given by equations (9) and (12) from Z\&Z.

The ``$C_1$'' term in Eq.~(\ref{eq:lin_C1}), 
which has a linear dependency on $\epsilon$, is termed as the mode-tide coupling strength (MTCS) by Z\&Z. This term is dominant for moderately high-order modes coupled with the tide. We use the same definition for MTCS and present our results in Sec.~\ref{sec:results}. 
The effects of the relativistic corrections on other terms with nonlinear dependency on $\epsilon$ for higher-order p and g-modes are discussed briefly in the appendix~\ref{sec:app_higher_order}.

\section{Mode-Tide Coupling Strength Results}
\label{sec:results}

\subsection{Newtonian versus Relativistic Equilibria (non-static tides)}


Before applying the relativistic corrections on the MTCS computation, we investigate how important relativistic corrections are. For this purpose, we devise a test in which the mode-tide coupling strength is computed for a Newtonian star in hydrostatic equilibrium as the background (similar to the Lane-Emden model, but in general form to take any equation of state),  within a fully Newtonian formalism. In order to observe the maximum effect, we choose our softest EOS, SFHo, for this test. 
We adjust the central density to have a $1.4M_{\odot}$ star in our Newtonian star model (equal to the gravitational mass of our TOV models), and the gravitational potential is given by the Poisson equation.  
We find that the eigenfrequency of the g-mode (n=32, l=4) for this model is $\approx 9.207$Hz.

\begin{figure}
  \includegraphics[width=\linewidth]{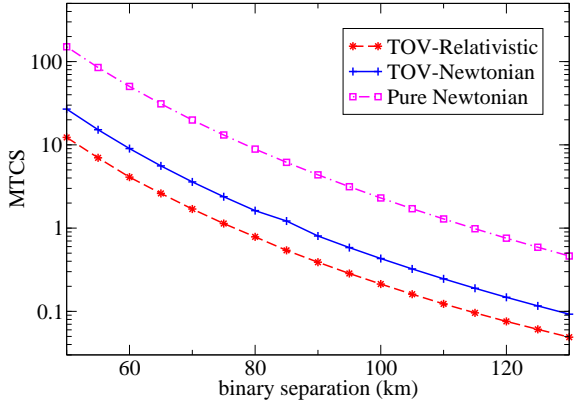}
  \caption[]{ 
MTCS versus binary separations computed for SFHo. Comparing the pure Newtonian MTCS computed for a Newtonian star with the hybrid methods: 1-TOV star and Newtonian MTCS as given by Z\&Z (the blue curve) , 2-TOV star with relativistic corrections for MTCS (the red curve).   
    }
  \label{fig:Lane-Emden-MTCS}
\end{figure}

The results of this test are displayed in Fig.~\ref{fig:Lane-Emden-MTCS}. 
In this figure, `Pure Newtonian' curve presents the modes and MTCS computed within the Newtonian formalism with a Newtonian star as the background in equilibrium, while the `TOV-Newtonian' curve shows MTCS for the hybrid method used by Z\&Z, i.e., Newtonian mode and MTCS computation with a TOV star as the background, and finally we have `TOV-Relativistic', which presents our hybrid method: the same TOV equilibrium, relativistic mode calculation, and Newtonian MTCS computation. 
We use Eqs.~(43)-(45) for non-static tides from Z\&Z for this purpose.

The MTCS computed for the Newtonian star is larger by about one order of magnitude comparing to the hybrid cases with a TOV star. Therefore, for a pure Newtonian model the nonlinear coupling instability can be triggered in early inspiral even for a moderately high-order mode. This also shows that the Z\&Z's hybrid method stands between the pure Newtonian model and our hybrid method with the relativistic corrections.  However, this test suggests that for more realistic and consistent results one should derive the mode-tide coupling coefficient in a fully relativistic formalism.

\subsection{Non-static Tides}

We follow Z\&Z's computation for the dynamical tide in Newtonian physics (see Eqs.~(43-45) in Z\&Z), and redo each calculation by including general relativistic corrections. The MTCSs are computed for a wide range of binary separations, 50-130km, for the $n=32$ and $l_g=4$ g-mode.

\subsubsection{Relativistic effects}
\label{subsubsection:rel-effect}

The relativistically-corrected MTCS values are compared with the Newtonian MTCS in Figs.~\ref{fig:shen-MTCS}, ~\ref{fig:SFHo-MTCS} and  \ref{fig:DD2-MTCS}.
Generally, the relativistic corrections do not introduce dramatic changes in the mode-tide coupling strength, especially for stiffer EOS. For the Shen EOS, the relativistic corrections suppress the nonlinear coupling strength by only 5-10\%. On the other hand, the relativistic terms become more important for more compact stars (i.e., softer equations of states), as one would expect: The coupling strength is suppressed by about 50\% for our softest EOS, namely, SFHo.  
DD2 appears to be an exception, where the MTCS rises up to 40\% after applying relativistic corrections. As we will discuss below, the DD2 case is rather special, the MTCS being extremely sensitive to the properties of the star near the surface.
The MTCS for DD2 models is found to be higher in the relativistic than in the Newtonian case for all of the treatments of the neutron star outer layers and surface that we have tried.


\subsubsection{Mode coupling instability}
\label{subsubsection:instability}

Following the instability analysis by Z\&Z, the shifted modal frequency (due to the coupling with the tide) becomes imaginary when the MTCS value exceeds 1, which causes the mode to grow exponentially with growth rate $\omega_g\sqrt{MTCS-1}$ by extracting energy from the orbital motion (see Sec. 4.4, and Eq.~(48) in Z\&Z). 

The binary separation at which the instability sets in is shown in table~\ref{tab:threshold} for Newtonian and relativistically-corrected  methods for the three EOSs. 
The suppressing effects of the relativistic corrections for SFHo make the mode-tide coupling (hydrodynamic) instability triggered only later in the inspiral phase. Even though the relativistic coupling is stronger for DD2, the MTCS still stays below 1 in our selected range of binary separations, as shown in Fig.~\ref{fig:Rel-MTCS-EOS}. In other words, for Shen and DD2, tidal disruption or plunge may occur before the onset of mode-tide instability, at least for this particular g-mode and 
unless more complete computations reveal a substantial increase in the separation at which the instability sets in.

\subsubsection{Surface issues and DD2}
\label{subsubsection:surf-dd2}

We found that the MTCS computation is very sensitive to the boundary defined by the minimum of the electron fraction (see Fig.~\ref{fig:Ye-EOS}). This is true for both static and non-static tides. This feature has been illustrated clearly in Fig.~10 of Z\&Z, as well, in which one notices that the largest contribution to the integral comes from the region very near the surface.  This issue is quite severe for the DD2 models we studied, where we observed high-frequency oscillations and sharp features close to the boundary. This boundary characteristic made our MTCS results unreasonably small at larger binary separation.

To investigate this issue more systematically, we carried out a few tests: First, we tested the accuracy of our numerical results by tightening the tolerance of the ODE solver by an order of magnitude, and found that the results remain unchanged. Second, we observed that the Brunt-V\"{a}is\"{a}l\"{a} frequency values change rapidly between grid points close to the boundary. To study its impact, we eliminated this feature by artificially smoothening $N^2$, and as a result the MTCS plot followed the same slope as the other equations of state when the binary separation is increased. Therefore, we focused on the buoyancy frequency and recomputed the MTCS with SFHo for several binary separations to study how much the results from other EOS depend on the smoothness of the buoyancy frequency. We find that the MTCS values get altered only by less than a percent with smoothed $N^2$; so the extreme sensitivity to the shape of $N^2$ vs $r$ near the surface is not found to affect every EOS.

For further examination, we computed MTCS using buoyancy frequencies that are smoothed in different ways, i.e., the smoothed value of $N^2$ for each grid point is given by taking average over a different number of grid points from the left and right sides, which makes the $N^2$ curve slightly different near the boundary.
This time the MTCS changed by $5-10\%$, proving that for DD2, the MTCS depends not only on the smoothness of $N^2$, but also on the shape of the buoyancy curve at the boundary.  
As a final test, we increased the number of the grid points in the original DD2 EOS table by a factor of 5 to increase the accuracy of the numerical derivatives used in the buoyancy frequency computation. Here the results remain almost equal to the original results with non-smoothed $N^2$. 

In conclusion, the MTCS is very sensitive to the shape and smoothness of $N^2$ near the surface for a neutron star modeled with the DD2 EOS. This should be considered as a physical feature of this EOS, since we have confirmed that the sensitivity is not mitigated by increasing the accuracy of the numerical integration of the eigenmode equations.  It indicates that, for this EOS, the MTCS is quite sensitive to behavior near the surface. This motivates future studies on handling this region carefully, particularly accounting for the crust, assuming it persists in this part of the late inspiral.

\subsubsection{Resonant couplings}
\label{subsubsection:res-coupling}

\begin{figure}
  \includegraphics[width=\linewidth]{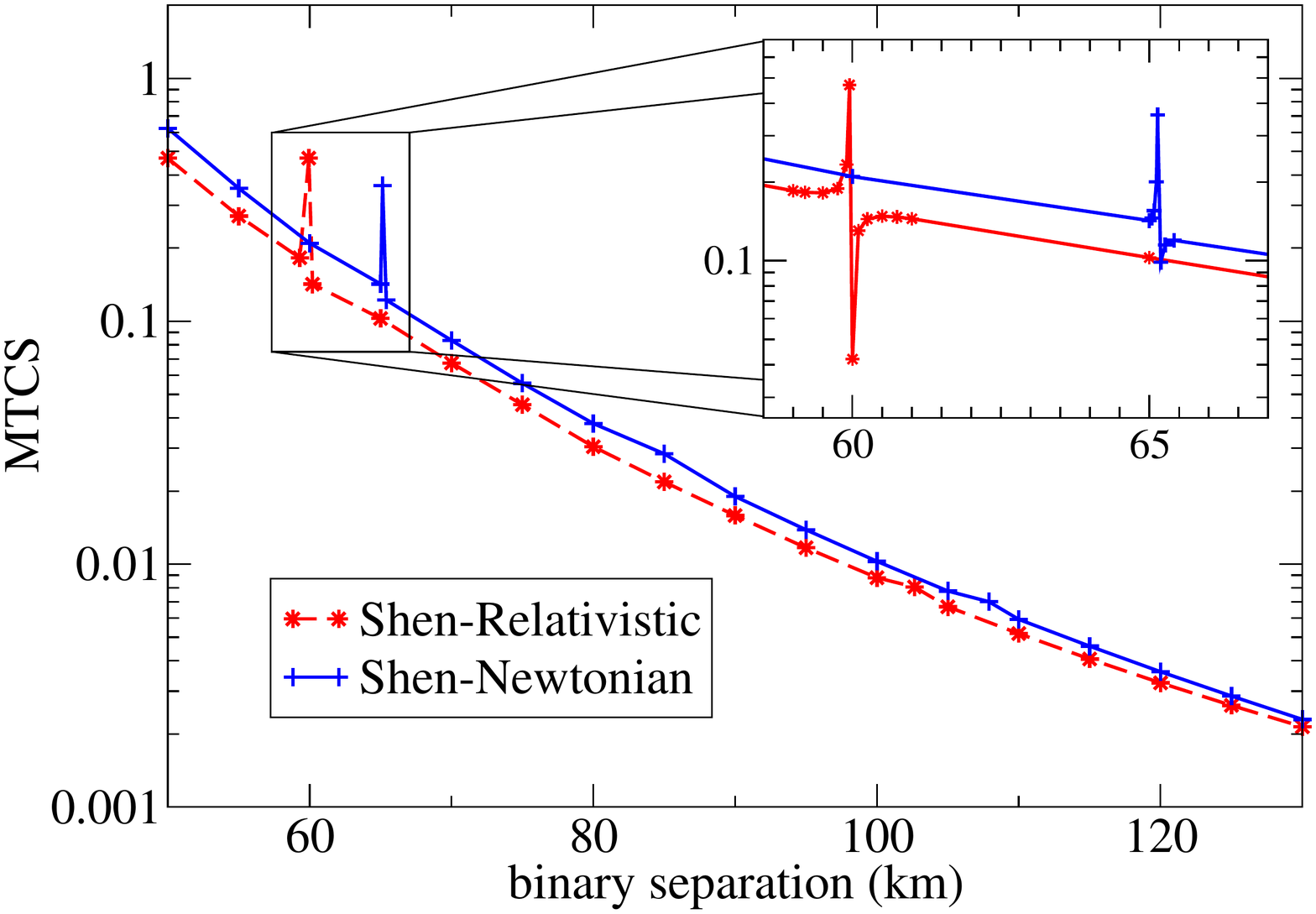}
  \caption[shen-MTCS]{ 
The Newtonian and relativistic MTCS computed for Shen EOS with different 
binary separation.
    }
  \label{fig:shen-MTCS}
\end{figure}

\begin{figure}
  \includegraphics[width=\linewidth]{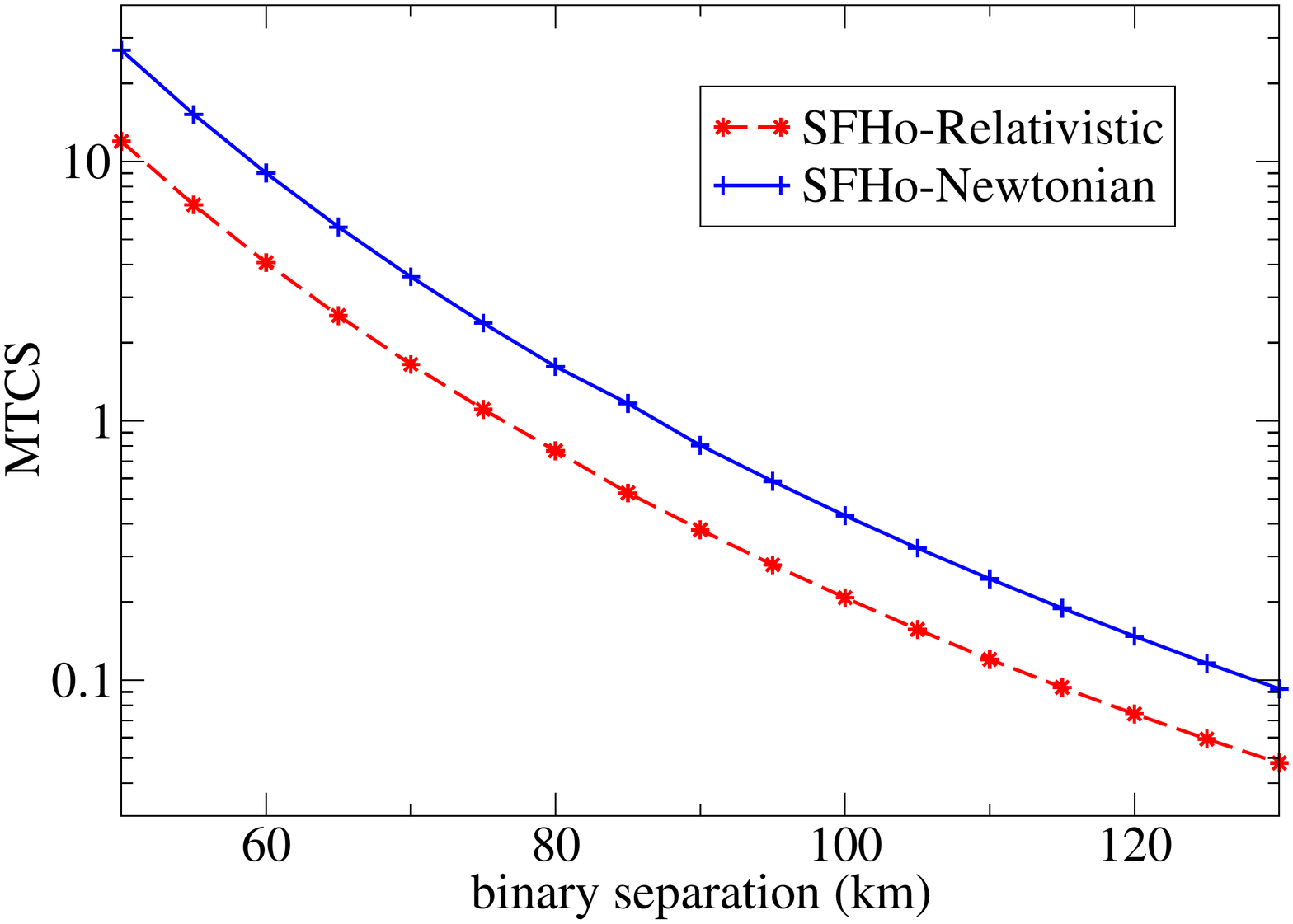}
  \caption[SFHo-MTCS]{ 
The Newtonian and relativistic MTCS computed for SFHo EOS with different 
binary separation.
    }
  \label{fig:SFHo-MTCS}
\end{figure}

\begin{figure}
  \includegraphics[width=\linewidth]{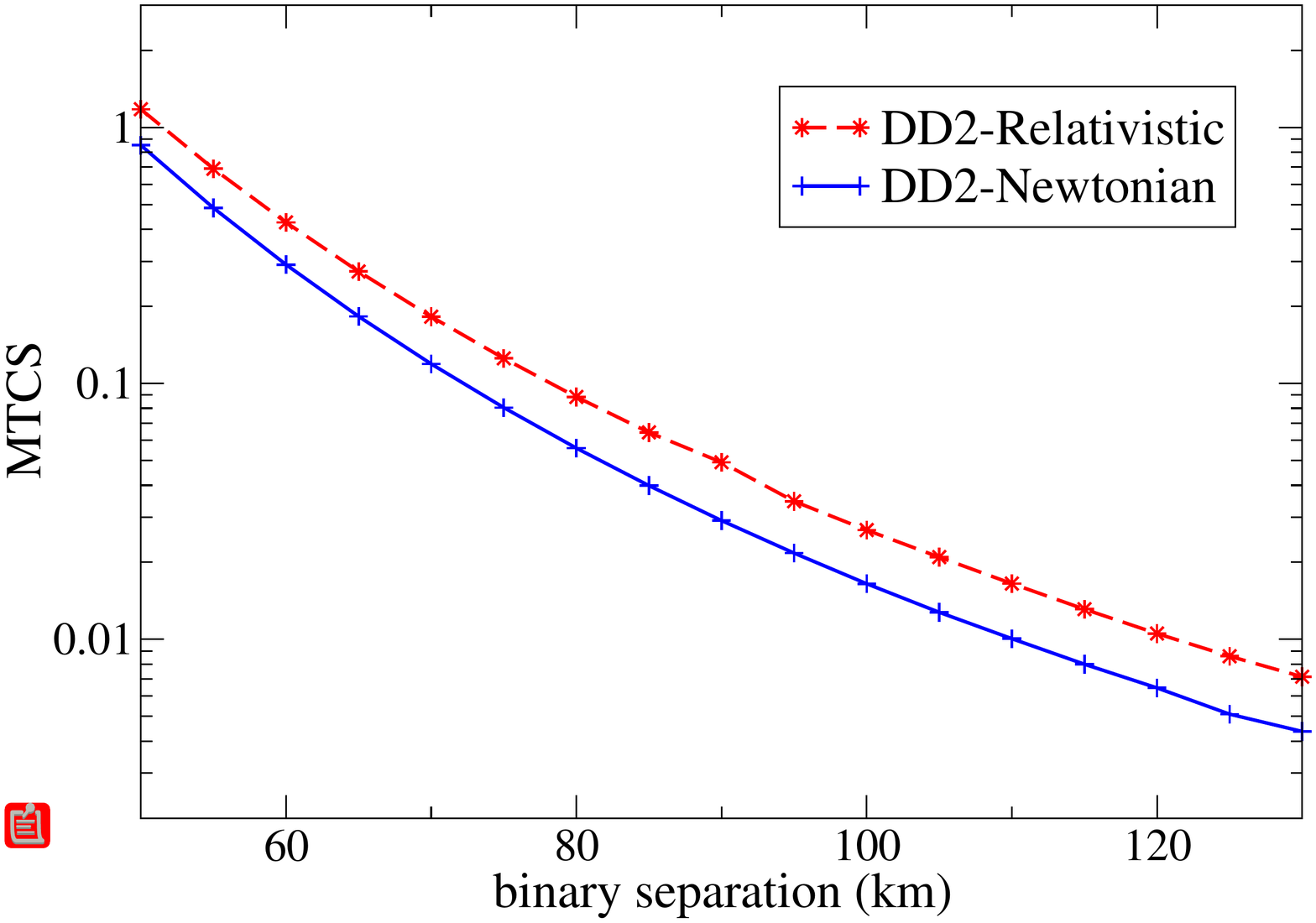}
  \caption[DD2-MTCS]{ 
The Newtonian and relativistic MTCS computed for DD2 EOS with different 
binary separation.
    }
  \label{fig:DD2-MTCS}
\end{figure}

When computing the MTCS, we observed that at certain binary separations the tidal displacement amplitude becomes extraordinarily large -- about an order of magnitude higher than the amplitude at neighboring separations.  This appears as distinguishable peaks in the MTCS plots, as shown in Fig.~\ref{fig:shen-MTCS} for our Shen case. Z\&Z observed similar peaks in their computations. Further investigation proves that this feature is caused by the linear resonant coupling of the dynamical tide with the $l=2$ g-modes (see Weinberg et al. (2012)~\cite{Weinberg_2012}). 
In other words, whenever the orbital frequency matches any of $l=2$ g-mode's frequencies they couple linearly to the dynamical tide, and this makes the terms with the tidal displacement dominant in the nonlinear MTCS computation. 
For instance, in Fig.~\ref{fig:shen-MTCS}, the sharp peak for the Newtonian result at $\approx65$km matches with the $l=2$, $n=5$ g-mode.

Resonant couplings in binary neutron stars have been studied in great detail~\cite{Lai:1993di,Lai_2017}.  The very high amplitude of the tidal displacement is partly an artifact of ignoring the inspiral during the calculation of eigenmodes.  In reality, the star will only be resonantly excited for a finite time.  (See, e.g., Ref.~\cite{Lai:1993di} for an amplitude calculation that accounts for the inspiral.)

Obviously, one expects to see similar peaks for other cases.  However the spacing we chose for the binary separations is not tight enough to capture all the possible resonant peaks. We also found that some of the resonant peaks occuring at some binary separations, such as $r\approx102$km and $r\approx113$km in Fig.~\ref{fig:shen-MTCS}, are smaller and less noticeable.  

\subsubsection{EOS comparison}
\label{subsubsection:eos-compare}

\begin{figure}
  \includegraphics[width=\linewidth]{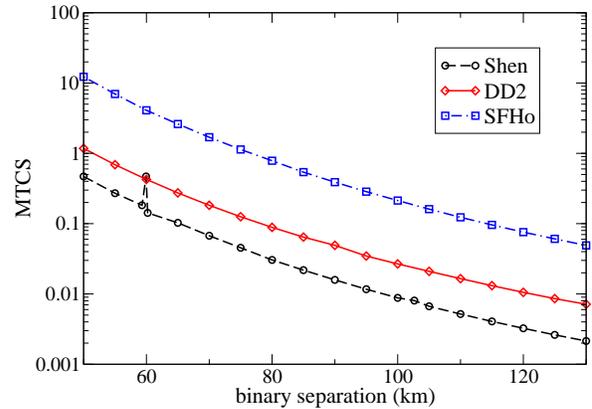}
  \caption[Rel-MTCS-EOS]{ 
The relativistic MTCS computed for Shen, DD2 and SFHo with different 
binary separation.
    }
  \label{fig:Rel-MTCS-EOS}
\end{figure}

Our studies confirm Z\&Z's conclusion on the dependence of MTCS on the EOS, which continues to hold even after applying the relativistic corrections. Similarly, our results confirm that the stiffness of the EOS and the buoyancy frequency affect the MTCS simultaneously (compare our Fig.~\ref{fig:Rel-MTCS-EOS} with Fig.~12 from Z\&Z). 
For instance, the MTCS for SFHo is bigger than the other two EOS by an order of magnitude. This is similar to Z\&Z's results for Sly4 compared to their other EOS.
It is clear that the EOS with the smallest buoyancy frequency yields the strongest mode-tide couplings.
However, the number of EOSs in our study is not large enough to support more general and detailed conclusions regarding EOS effects.

\begin{table}
\begin{ruledtabular}
\begin{tabular}{ l l l l l }
EOS & $f_{g(N)} (Hz)$ & $f_{g(R)} (Hz)$ & $MTCS_{(N)}$ & $MTCS_{(R)}$ \\
\toprule
Shen & 36.06 & 27.505 & 0.08339 & 0.06616  \\
DD2 &  14.15 & 11.352 & 0.05385 & 0.06404  \\
SFHo & 11.45 & 8.817  & 3.5891 & 1.6752  \\
\end{tabular}
\end{ruledtabular}
\caption{\label{tab:MTCS_results}
The Newtonian and relativistic eigenfrequencies with the Cowling approximation, and MTCS evaluated for g-mode n=32, l=4, for dynamical tides with binary separation $A=70$km.
The Newtonian eigenfrequencies correspond to the solution of the Newtonian perturbation equations with TOV background stars in equilibrium. \\
}
\end{table}

\begin{table}
\begin{ruledtabular}
\begin{tabular}{ l l l }
EOS & ${\rm Threshold}_{(N)}$km & ${\rm Threshold}_{(R)}$km \\
\toprule
Shen & 46.14 & 43.97 \\
DD2 & 41.37  & 45.01 \\
SFHo & 87.38  & 75.77 \\
\end{tabular}
\end{ruledtabular}
\caption{\label{tab:threshold}
The Newtonian and relativistic instability thresholds, ${\rm Threshold}_{(N)}$ and ${\rm Threshold}_{(R)}$, respectively. These are the binary separations where the MTCS reaches unity.\\
}
\end{table}


\subsection{Static Tides}

The computations of MTCS for static tides have been performed in Z\&Z for different EOS using a novel technique called the volume preserving transformation (VPT) introduced by Venumadhav et al.~\cite{Venumadhav_2013}. This transformation maps a tidally deformed star into a radially stretched spherical star of equal volume. Here we follow the same approach and compute our Newtonian MTCS using Eq.~(33) from Z\&Z. However, for our relativistic corrections for MTCS, we only consider the relativistic internal modes and we leave the tidal displacement Newtonian to be able to use the same VPT technique. This computation is only done to examine how effective the relativistic corrections are for the static tide MTCS.

The MTCS computed for the static tide confirms the main results in Z\&Z. As it has been claimed in Venumadhav et al.~\cite{Venumadhav_2013}, the four-mode coupling cancels 
the three-mode coupling for the static tide, and makes the mode-tide coupling instability unimportant for this case.
Our results for Shen, DD2 and SFHo 
show that relativistic correction applied on the g-mode computation increase the mode-tide coupling strength only by ten to thirty percent relative to that in the Newtonian case.
However, the MTCS is still too small (MTCS$\ll 1.0$) to trigger the instability even when the neutron stars are extremely close (i.e., about 35km apart) before the merger. 
The MTCS values for static tide are shown in table~\ref{tab:MTCS_static} for different EOS, for two different  binary separations:
a) $A=100$km, b) $A=2R$ (when the stars are touching). 
Obviously the $A=2R$ cases are not realistic, since the neutron stars would be disrupted earlier due to the tidal forces.

\begin{table}
\begin{ruledtabular}
\begin{tabular}{ l l l l }
EOS & Binary Separation & $MTCS_{(N)}$ & $MTCS_{(R)}$ \\
\toprule
Shen & 100 km  & $2.58\times 10^{-4}$  & $2.82\times 10^{-4}$ \\
DD2 &  100 km  & $3.49\times 10^{-4}$ & $4.18\times 10^{-4}$ \\
SFHo & 100 km  & $3.12\times 10^{-4}$  & $3.35\times 10^{-4}$ \\
\hline
Shen & 2R  & 0.2466  & 0.2621 \\
DD2 &  2R  & 0.2955  & 0.3901 \\
SFHo & 2R  & 1.07    & 1.19    \\
\end{tabular}
\end{ruledtabular}
\caption{\label{tab:MTCS_static}
The Newtonian and relativistic MTCS for static tide evaluated for g-mode n=32, l=4,
with binary separations $A=100$km and $A=2R$.\\
}
\end{table}

\section{Conclusion}
\label{sec:conclusion}

In this paper, we have studied the importance of relativistic corrections in the nonlinear mode-tide coupling in neutron star binary systems. 
The background stars in the binary systems are assumed to be relativistic and identical.
We have found the MTCS for both static and non-static tides both in pure Newtonian physics and with modes and tides computed in general relativity. We perform this comparison for a collection of three three-nuclear-fluid tabulated EOS at minimum temperature and for a range of binary separations.  We mainly focus on non-static tides because of their importance in the instability analysis.

Although we calculate dynamical tides and stellar eigenmodes in general relativity, the MTCS integral is still in the Newtonian formalism, originally derived by Weinberg et al. (2012)~\cite{Weinberg_2012}.  Thus, even our relativistic treatment is not fully relativistic.  Nevertheless, it provides a strong indication of how important relativistic effects will be for a given binary separation and neutron star compaction.

Our results show the relativistic corrections for non-static tides make a small difference in the MTCS values for the stiffest EOS, but more significant changes (up to 60\%) for a soft EOS such as SFHo. 
We have also investigated a few cases for static tides with the relativistic corrections, and they all show that the couplings are still too weak to be important for the instability analysis.
We also confirmed Z\&Z's argument on the dependence of the coupling strength on the EOS (i.e., EOSs with smaller buoyancy frequency have stronger couplings). 
We also observed that the accuracy of our computations is sensitive to the outer boundary of the MTCS integral (especially for DD2).

The following arguments are needed to be considered in future studies of the nonlinear mode-tide couplings for a better accuracy, 
1- To compute a fully relativistic MTCS, one needs to derive the coupling coefficient in a relativistic formalism.
2- The core of a cold neutron star is expected to be superfluid, and based on superfluid models with finite temperature, the bouyancy frequency is predicted to be much smaller than normal fluids (see Gusakov \& Kantor (2012)~\cite{Gusakov_2012}). This assumption can lead to significantly stronger couplings, since our results and similar studies show the MTCS is highly dependent on the buoyancy frequency. 
3- We found that the MTCS is highly sensitive to the location of the boundary and the computed quantities at the boundary, such as buoyancy frequency. Future studies will need a better treatment for the core-crust boundary, and also utilize more realistic models (e.g., solid crusts for neutron stars).
4- For a complete set of studies, specifically to investigate the p-g instability, one needs to consider extremely higher-order g-modes coupled with dynamical tide and all p-modes for different equations of state, including all the terms with non-linear dependency on $\epsilon$ for shifted mode's calculations in eq.~\ref{eq:freq_shift}.

\acknowledgments

The authors thank Nils Andersson, Pantelis Pnigouras, Dong Lai and Yixiao Zhou for helpful discussions and advice over the course of this project. We would also like to thank Nathan Johnson-McDaniel for thoroughly reading the manuscript and making several useful comments.
We also thank the anonymous referee for comments and suggestions that helped us to improve our manuscript.
F.H. acknowledges support
from the Navajbai Ratan Tata Trust and LIGO-India at IUCAA, India, and grant No. 2019/35/B/ST9/04000 from the Polish National Science Center, Poland.  M.D. gratefully acknowledges support from the NSF through grant PHY-2110287.

\appendix
\section{Numerical Methods and Tests}
\label{sec:app_methods}

\subsection{Shooting Method with Fitting Point}

We numerically integrate Eqs.~(\ref{eq:H1}),  (\ref{eq:K}), (\ref{eq:W}) and (\ref{eq:X}), for dynamical tidal displacement, 
and we integrate Eq.~(\ref{eq:relxi}) and (\ref{eq:relp}) for eigenmodes.
We solve these systems of ODEs by applying the boundary conditions at the center and surface of the star using a shooting method. The boundary conditions are discussed in~\ref{subsec:tide_dis} and~\ref{subsec:eigenmodes}.
We integrate the equations outward from the center and inward from the surface to a fitting point in the middle of the star.  A multidimensional root finder is used to match the solutions at this point.  In particular, we find the roots for $W$ and $H_1$ at the center, and $W$ at the surface for the tidal displacement. For the eigenmodes on the other hand, the root finder gives us the frequency and $\xi_r$ at the surface.
We use several routines from Numerical Recipes ~\cite{NumRec} including the 4th order Runge-Kutta ODE solver and Broyden's multidimensional root finder.  

\subsection{Tests: Newtonian and Relativistic Eigenmodes with the Cowling approximation}

To compute the stellar oscillation eigenmodes in relativistic and Newtonian limits, both with the Cowling approximation, we developed our own code (the results are given in~\ref{subsec:eigenmodes}).

To derive the eigenmodes, we solve the linear perturbation equations with the Cowling approximation, namely, Eqs.~(\ref{eq:relxi}) and (\ref{eq:relp}). 
We start with an initial guess to bracket the f-mode frequency. From the literature we know that the f-mode frequencies for neutron stars are of the order of a few kHz. With some trial-and-error, we find the f-mode frequency with zero nodes along the radius of the star. The code automatically adjusts the brackets for higher frequencies to generate results for p-modes, and lower frequencies for g-modes. A separate routine counts the number of the nodes along the radius of the star for each eigenfunction output by the ODE solver to determine the radial number of the mode. We continue running the code till we find the desired $n$ (radial number) eigenmodes.

These codes were successfully tested by computing the eigenfrequencies and eigenfunctions of a series of 
Lane-Emden stars (Robe 1968)~\cite{Robe_1968} for the Newtonian code,
and TOV models (Font et al. 1999)~\cite{Font_2000} for the relativistic code. 
All these models have polytropic equations of state with polytropic index
$n=2$ for the Newtonian comparison and 
$n=1$ for the relativistic comparison.
Tables~\ref{tab:eigen-newt-poly} and ~\ref{tab:eigen-rel-cowling-poly} show the eigenfrequencies computed with our codes against 
the result from ~\cite{Robe_1968} and ~\cite{Font_2000}. Our results agree with theirs up to $\approx 0.1\%$.

\begin{table}
\begin{ruledtabular}
\begin{tabular}{ l l l l }
Mode & $\omega^2 $(Newtonian code) & $\omega^2 $(Robe-1968) \\
\toprule
p6 &  157.7  &  157.8  \\
p5 &  119.7  &  119.8  \\
p4 &  86.75  &  86.77  \\
p3 &  58.84  &  58.85  \\
p2 &  36.02  &  36.02  \\
p1 &  18.39  &  18.39  \\
\hline
f  &  6.080  &  6.074  \\
\hline
g1 &  0.7772 &  0.7761  \\  
g2 &  0.4025 &    -     \\ 
g3 &  0.2479 &  0.2473  \\ 
g4 &  0.1686 &  0.1682  \\ 
g5 &  0.1224 &  0.1220  \\
g6 &  0.09296 & 0.09265  \\
\end{tabular}
\end{ruledtabular}
\caption{\label{tab:eigen-newt-poly}
The eigenfrequencies of a Lane-Emden star, with polytropic EOS, and
polytrope index $n=2$, $\gamma=1+1/n=5/3$ and $l=2$. 
The unit of $\omega^2$ is $\pi  \rho_\text{avg}$, 
where $\rho_\text{avg}$ is the average density. 
The unit of $\omega$ conforms to that used by Robe-1968. Note that Robe-1968 did not list the  frequency of $g_2$.   \\
}
\end{table}

\begin{table}
\begin{ruledtabular}
\begin{tabular}{ l l l l }
Mode & Relativistic Cowling (kHz) & Font et al.(1999) (kHz) \\
\hline
f &  1.8825 &  1.8843 \\
p1 &  4.1060 &  4.1099 \\
p2 &  6.0298 &  6.0351 \\ 
p3 &  7.8670 &  7.8733 \\
p4 &  9.6663 &  9.6740 \\
\end{tabular}
\end{ruledtabular}
\caption{\label{tab:eigen-rel-cowling-poly}
The eigenfrequencies of a TOV star, within the Cowling approximation, and with polytropic EOS,
polytrope index $n=1$, $\kappa=100$ and $\rho_{0c}=0.00128$,
in the polytropic EOS: $P=\kappa \rho_0^{1+1/n}$
}
\end{table}

\subsection{Tests: Relativistic fundamental mode with space-time perturbations for a single NS}

To test our solutions for relativistic perturbation equations without the Cowling approximation (which is used to derive the tidal displacement in ~\ref{subsec:tide_dis}), we have carried out the following tests: 1) computing the f-mode frequencies of a few perturbed TOV models and comparing them with the literature to test our code in deriving the fundamental modes without tides. 
2) computing the tidal Love number for static tides, and comparing the numerical exterior solution with the known analytical solution, which is explained in the next section.

For the first set of tests, we reproduced the results from 
Chirenti et al.(2012)~\cite{Chirenti_2012} and (2015)~\cite{Chirenti_2015}
for a perturbed single neutron star. 
We solve the coupled ODEs Eq.~(\ref{eq:H1}),   (\ref{eq:K}), (\ref{eq:W}) and (\ref{eq:X}) for interior as it is explained in Sec.~\ref{subsec:tide_dis}. 

The solution outside of the star is 
derived from the Zerilli Eq.~(\ref{eq:zerilli}), by applying the outgoing wave as
the boundary condition at infinity (see Eqs.~(A35-A38) from Lindblom \& Detweiler~(1983)~\cite{Lindblom:1983ps}).

\begin{figure}
  \includegraphics[width=\linewidth]{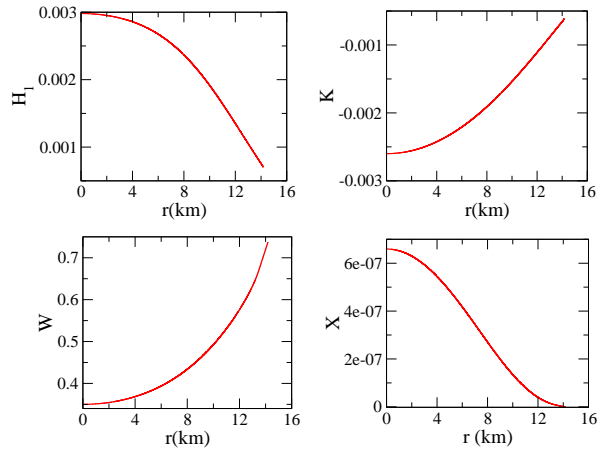}
  \caption[eigen-poly-inside]{ 
The hydrodynamic and metric perturbation functions inside the star,
for a polytrope model with $\Gamma=2$.
    }
  \label{fig:eigen-poly-inside}
\end{figure}

\begin{figure}
  \includegraphics[width=\linewidth]{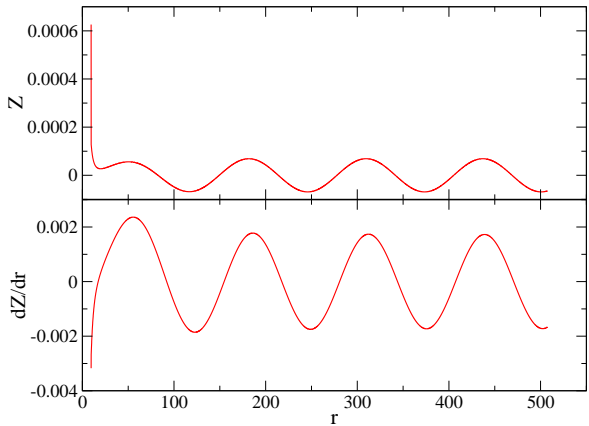}
  \caption[Zerilli-poly]{ 
Metric perturbations outside the star, given in terms of the Zerilli function Z and its derivative.
    }
  \label{fig:Zerilli-poly}
\end{figure}

The eigenfunctions $H_1$, $K$, $W$ and $\chi$ for the interior solutions
and the Zerilli function for the exterior solution are shown in Figures~\ref{fig:eigen-poly-inside} and~\ref{fig:Zerilli-poly} for a polytropic 
TOV star with polytropic index $=1$, and polytropic constant $\kappa = 100$. 
These results are in good agreement with figures (1) and (2) from Chirenti~{\it et al.}~(2012)~\cite{Chirenti_2012}.
The f-modes frequency we derive for this polytrope model is $1.5711$ kHz, which if off by only $ \approx 0.51 \% $ comparing with~\cite{Chirenti_2012}.
The f-mode frequencies for neutron stars with LS220  
EOS and different masses are presented in table~\ref{tab:f-mode-test} 
for comparison with Chirenti~{\it et al.} (2015)~\cite{Chirenti_2015}.  

\begin{table}
\begin{ruledtabular}
\begin{tabular}{ l l l }
Mass ($M_{\odot}$) & f (kHz) & f (kHz) (as in Ref.~\cite{Chirenti_2012}) \\
\hline
2.053  & 2.465  & 2.452  \\ 
1.788  & 1.984  & 1.969  \\
1.525  & 1.782  & 1.777  \\
1.273  & 1.618  & 1.628  \\
\end{tabular}
\end{ruledtabular}
\caption{\label{tab:f-mode-test}
The f-mode frequency of a TOV star within the full relativistic formalism for LS220
at several neutron star masses. For comparison we also list the corresponding values from Chirenti et~al.~\cite{Chirenti_2012} in the right most column.\\
}
\end{table}


\subsection{Test: Metric perturbation for Static Tides}

\begin{figure}
  \includegraphics[width=\linewidth]{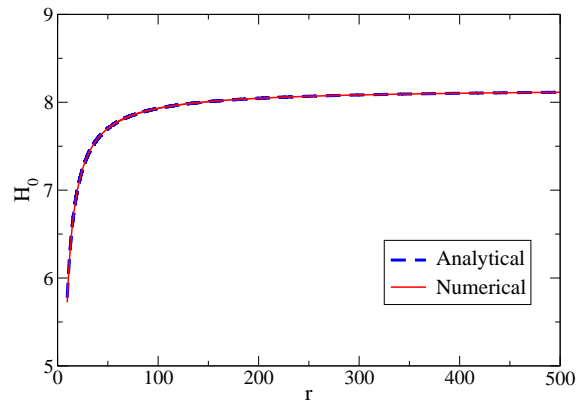}
  \caption[H0-static-test]{ 
$H_0$ metric perturbation function out side the star comparing the 
analytical and numerical solutions.
    }
  \label{fig:H0-static-test}
\end{figure}

We compare our numerical exterior solution with the analytical solution given by Eq.~(21) from Ref.~\cite{Hinderer_2008}.
We study the static tides for this test and, therefore, set $\omega = 0$ in our equations. For exterior, we use the Zerilli equation integrator from Sec.~\ref{subsec:tide_dis} to find the Zerilli function 
to derive $K$ and $H_1$ by matching to the asymptotic solution at large radii given by~\cite{Hinderer_2008} Eq.~(21).
We use these perturbation functions to derive $H_0$ at the surface using Eq.~(\ref{eq:H0}).
$H_0$ is used in Eq.~(23) from ~\cite{Hinderer_2008} to compute the tidal Love number. 
It is worth mentioning that the definitions of $H_0$ and $K$ in 
Hinderer~(2008)~\cite{Hinderer_2008} are different by factor of $r^l$ from those in Detweiler and Lindblom~(1985)~\cite{Detweiler_1985}.
Figure~\ref{fig:H0-static-test} compares the numerical and analytical exterior solutions for $H_0$ function, and shows good agreement between the two.

\section{Relativistic corrections for higher-order modes}
\label{sec:app_higher_order}

The problem of the nonlinear three-mode coupling has been studied by Weinberg et al.~\cite{Weinberg_2012} in a binary system for an arbitrary parent mode and a pair of daughter modes. However, for a neutron star binary system, only the couplings between the dynamical tidal field (parent mode) and higher-order daughter modes are strong enough to introduce the instability and leave measurable imprints in gravitational waves from the inspiral phase~\cite{Weinberg_2016}.
In addition, for a complete studies of the p-g instability for higher-order modes, it is necessary to include the four-mode coupling terms, as well as three-mode coupling terms with non-linear dependency on the tidal strength (as proven by Venumadhav et al.~\cite{Venumadhav_2013} and mentioned in Sec.~\ref{subsec:MTCS-and-pgi}).
In this appendix, we try to investigate the effects of the relativistic corrections only on the three-mode coupling term for higher-order modes, and only for one particular case of a p and g-mode pair. We postpone the complete calculations to the future studies.

To investigate how large the mode-coupling strength is for such cases, and how much difference appears by applying the relativistic corrections, we consider two cases; 
First, the nonlinear coupling between dynamical tide and $n=94$ g-mode, and second, the nonlinear coupling of dynamical tide and a pair of $n=97$ p-mode and $n=94$ g-mode.
These two modes have similar wavelengths in the NS interior and a correspondingly significant overlap of eigenfunctions, (see Fig.(3) from~\cite{Weinberg_2012} illustrating an overlapping p-mode and g-mode pair.  
Here we perform two types of comparisons between the Newtonian and relativistic results:
a) the coupling between the g-mode and the tide, which is linear on the tidal strength, and b) the three-mode coupling between the tide, a g-mode, and a p-mode, which is quadratic in the tidal strength.
The former is the same MTCS value that we calculated for the results section~\ref{sec:results}, while the latter is a new series of calculations done only for high-order mode cases.
Our studies for this part are limited to the SFHo EOS to investigate the maximum likely effect of the relativistic terms. The Newtonian and relativistic frequencies of the selected p-mode and g-mode in the Cowling approximation are presented in Table~\ref{tab:high-p-g}.

Referring to Eq.~(\ref{eq:freq_shift}), the first comparison is done for the ``$C_1$'' term (or equivalently MTCS). We call it the ``linear'' term, because of the linearity in tidal strength $\epsilon$; see Eq.~(\ref{eq:lin_C1}).
The second comparison is done only for the ``$C_3$'' term, which is the order of $\epsilon^2$ (see Eq.~(\ref{eq:nonlin_C3})), so we label it as the ``nonlinear'' term.
Similar to the linear term, the nonlinear term contains the homogeneous and inhomogeneous parts, which are derived from Eqs.~(A55-A62) and Eq.~(A72) from~\cite{Weinberg_2012} (see section 4.2 in Z\&Z for more details).  
As it is mentioned in Z\&Z the nonlinear terms are expected to become dominant for higher-order mode-tide couplings in frequency shift calculations.

\begin{figure}
\includegraphics[width=\linewidth]{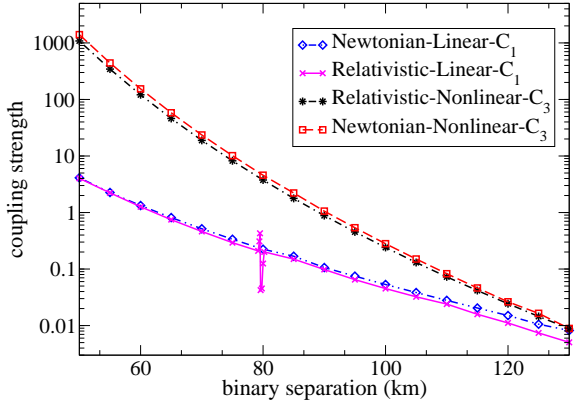}
  \caption[mtcs-g94-p97]{ 
The Newtonian and relativistic coupling strength computed for the linear  ($C_1$) and the nonlinear  ($C_3$) terms for the tide, p97 and g94 modes, ($l_p=l_g=4$), with the SFHo EOS. Here, p97 and g94 denote the p-mode of order n=97 and g-mode of order n=94, respectively.}
  \label{fig:mtcs-g94-p97}
\end{figure}

The comparison between the Newtonian and relativistic coupling strength, for linear and nonlinear terms, for high-order modes is presented in Fig.~\ref{fig:mtcs-g94-p97}. We, again, observe that our relativistic corrections cause suppressing effects, though the changes are very small. This shows that for high-order modes, the mode structure and coupling are more local and less affected by space-time curvature. 

At this point, it is interesting to note the changes of the linear term (MTCS) at higher-order modes. For this purpose one can compare the MTCS for SFHO, g32 case in Fig.~\ref{fig:Rel-MTCS-EOS} with the linear term (MTCS) for g94 case with the same equation of state in Fig.~\ref{fig:mtcs-g94-p97}. These results show that the MTCS does not change significantly as we go to the higher-order modes. In fact, this particular case shows that the MTCS decreases slightly for a relatively higher-order g-mode, with n=94 nodes.

It is important to mention that we observe sensitivity of the linear term to the surface region for the tide-g94 case in Fig.~\ref{fig:mtcs-g94-p97}. Similar to our DD2 case with tide-g32, the MTCS starts decreasing significantly as one goes to larger binary separations. This feature can be eliminated partially by applying a smoothed Brunt-V\"{a}is\"{a}l\"{a} frequency. Since in this paper our focus is on studying the relativistic effects on MTCS, we postpone further investigation of these boundary issues to future studies.   

In Fig.~\ref{fig:mtcs-g94-p97}, a peak is visible for the Newtonian linear term due to the mode-tide resonant coupling,
similar to the Shen's MTCS in Fig.~\ref{fig:shen-MTCS}.
However the peak is not observed for the Newtonian nonlinear term at the same binary separation because our selected high-order p-mode and g-mode pair can not satisfy the condition for the three-mode resonant coupling: $\omega_b + \omega_c \simeq \omega_a$.

Finally, it is worth mentioning that although our analysis is not complete in the calculation of the frequency shift for higher-order modes, nevertheless it shows that the relativistic corrections do not create a significant difference for C1 and C3 terms. Therefore, the Newtonian approach can be considered accurate enough for such cases.

\begin{table}
\begin{ruledtabular}
\begin{tabular}{ l l l }
mode & $f_{(N)}$ (Hz) & $f_{(R)}$ (Hz) \\
\hline                                         
$g_{94}$ & 4.0987 &  3.1193 \\
$p_{97}$  & $337.87 \times 10^{3}$ & $205.5 \times 10^{3}$ \\ 
\end{tabular}
\end{ruledtabular}
\caption{\label{tab:high-p-g}
The frequencies of high-order p-modes and g-modes computed in the Cowling approximation for TOV model with SFHo EOS, used for the coupling strengths computations. The subscript on the mode type is the radial mode number $n$; for all cases, $l_p = 4 = l_g$ above.\\
}
\end{table}


\bibliography{references.bib}
\end{document}